\documentclass[showpacs,prl,letterpaper,twocolumn,floatfix]{revtex4}

\usepackage{amsmath}
\usepackage{graphicx}
\usepackage{color}
\usepackage{calc} % math with lengths
\usepackage{xspace} % For spacing after units macros

\newcommand{\nm}{\ensuremath{\mathrm{\,nm}}\xspace}
\newcommand{\um}{\ensuremath{\,\mu\mathrm m}\xspace}
\newcommand{\mm}{\ensuremath{\mathrm{\,mm}}\xspace}
\newcommand{\THz}{\ensuremath{\mathrm{\,THz}}\xspace}

\newcommand{\G}{\overleftrightarrow{\mathbf G}}
\newcommand{\e}{\mathbf e}
\newcommand{\id}{\mathrm d}
\newcommand{\R}{\mathbf r}

\makeatletter
\renewcommand{\section}[1]{ % usually we wouldn't pass in the section as an argument but we need to in order to put the period ndash after it.
	\@startsection{section}{1}{\parindent}{0px}{0px}{\textit}{#1}.--
}
\makeatother

\newcommand{\rev}[1]{{\color{black}#1}}

\newlength{\figwidth}
\begin{document}
% title (fold)
\title{Disorder-induced coherent scattering in slow-light photonic crystal waveguides}

\author{M. Patterson}
\author{S. Hughes}
\email{shughes@physics.queensu.ca}
\affiliation{Department of Physics, Queen's University, Kingston, ON K7L 3N6, Canada}

\author{S. Combri\'e}
\author{N.-V.-Quynh Tran}
\author{A. De Rossi}
\email{alfredo.derossi@thalesgroup.com}
\affiliation{Thales Research and Technology, Route D\'epartementale 128, 91767 Palaiseau CEDEX, France}
\author{R. Gabet}
\author{Y. Jaou\"en}
\affiliation{Telecom ParisTech, 46 Rue Barrault, 75634 Paris CEDEX 13, France}

\date{\today}

\begin{abstract}
% We present a new theoretical formalism and matching experiments to describe disorder-induced coherent scattering in photonic crystal waveguides. Employing a self-consistent optical scattering theory, with only statistical functions to describe the structural disorder, we directly compare to transmission measurements and frequency-delay reflectometry maps obtained for high-quality GaAs photonic crystal membranes. The excellent qualitative agreement between theory and experiment provides clear physical insight into naturally occurring light localization and multiple coherent-scattering phenomena in slow-light waveguides.

\rev{We present light transmission measurements
and frequency-delay reflectometry maps for
%high-quality
GaAs photonic crystal membranes,
which show
% and demonstrate
  the transition from propagation with a well defined group
velocity to a regime  completely dominated by
disorder-induced coherent scattering.
Employing a  self-consistent optical scattering theory,
% for the structural disorder,
%that includes multiple forward and backward scattering events,
with only statistical functions to describe the structural disorder,
we obtain an excellent agreement with the experiments using no fitting parameters.
Our experiments and theory together
%The excellent qualitative agreement
provide clear physical insight into naturally occurring light localization and multiple coherent-scattering phenomena in slow-light waveguides.}

\end{abstract}
\maketitle

%title(end)

\section{Introduction} % (fold)
\begin{figure}
	\centering
	\includegraphics{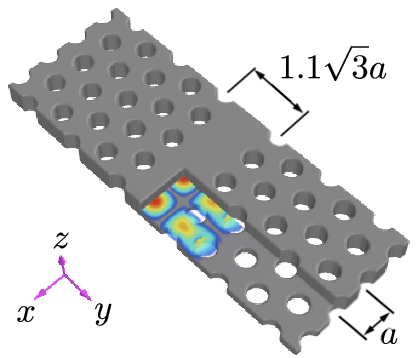} %4.3 wide x 3.6 cm tall
	\hspace{\stretch{1}}
	\includegraphics{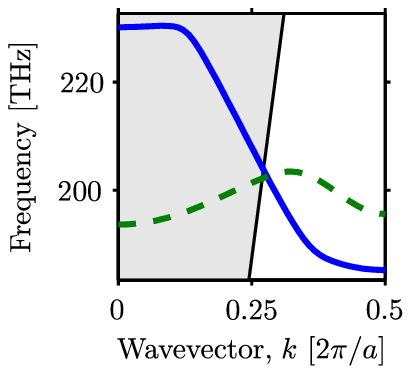} %[height=4.3cm,width=3.375in-4.4cm]
	\caption[]{\label{fig:structure}Left: \rev{Schematic of the planar PC waveguide. A section of the structure is cut away to show the anti-nodes in the $y$ component of the Bloch mode electric field.}
		Right: The computed corresponding band structure. The dispersion relations for the two lowest-order waveguide modes are shown and the lowest-order (blue, solid) mode is used for the calculations. The grey region at the left represents the continuum of radiation modes above the light line.}
\vspace{-0.4cm}
\end{figure}

Photonic crystals (PCs) are periodic dielectric structures that can strongly alter the propagation of light due to the interaction of coherent reflections from the constituent periodic features. In particular, semiconductor-based planar PCs (e.g.\ Fig.~\ref{fig:structure}) are of great interest due to their ease of fabrication using standard etching and lithography techniques. PC structures can exhibit interesting phenomena such as light trapping on sub-wavelength spatial dimensions in high-quality cavities~\cite{Akahane:2003}, or engineered waveguide band dispersions with a vanishing group velocity~\cite{Notomi:2001, Vlasov:2005, Baba:2008a}. Both of these effects give rise to novel regimes of light-matter interaction and have broad applications in nano- and quantum-technologies.

An ongoing challenge, is the theoretical description of real fabricated samples, including the generally unknown role of unavoidable structural imperfections, collectively termed ``disorder.'' Indeed, it is now well established that slow-light PC waveguides suffer from significant losses attributed to scattering at disordered surfaces and other device imperfections~\cite{Kuramochi:2005,Parini:2008}. Rigorous modeling of this generally undesired \emph{extrinsic} scattering phenomena is essential for understanding the underlying physics of measurements and for eventually producing functional devices. However, since the spatial scale of a disordered hole interface is only around 3\nm or less (for $\sim$200\nm holes), and because of the complexity of the
 structural disorder~\cite{Skorobogatiy:2005},
 %varies rapidly across the high-index-contrast interface~\cite{Skorobogatiy:2005},
 the theoretical description of light scattering presents enormous challenges.

Previously, \citet{Hughes:2005} introduced a PC wave\-guide model for \emph{incoherent} extrinsic scattering in the slow-light regime. \rev{The incoherent theory predicted the average scattering loss, using a first- and  second- order Born
approach for the scattering events, with the incoherent averaging performed over a large set of nominally identical samples (an ensemble average). The ``perturbative'' backscattering loss was predicted to scale with the group velocity $v_g$ as $1/v_g^2$ and dominate over scattering into radiation loss modes ($1/v_g$ scaling).} Similar scalings were also implied by \citet{Povinelli:2004} and by \citet{Gerace:2004}, though the scaling details depend on the nature of the propagating modes and
how they sample the disorder regions. These
%general
approximate loss-scaling relations have been confirmed experimentally by a number of groups, e.g.~\cite{Kuramochi:2005,OFaolain:2007,Engelen:2008}, \rev{but they may break down at low group velocities where multiple
disorder-induced scattering becomes significant.} Recent experimental measurements have reported interesting features such as narrow-band resonances near the band edge~\cite{Topolancik:2007,Parini:2008} that are not explained at all by the incoherent and perturbative scattering theories \rev{(see Figure~\ref{fig:spect})} and simple $v_g$ scaling rules.
% do not make sense.

\rev{In this Letter, we introduce a non-perturbative theory of \emph{coherent} optical scattering. In contrast to previous works, we include unlimited forward- and back-scattering events (through a coupled mode theory) which, combined with modeling individual, \emph{fully 3D}, disordered waveguides instead of ensemble averages (a statistical average
over many disordered waveguides), is essential to understand the new experimental observations.} The theoretical results are presented along side measurements on state-of-the-art GaAs PC waveguides which are probed with transmission measurements~\cite{Combrie:2006} and optical low-coherence reflectometry (OLCR). OLCR allows the measurement of the back-reflected signal as a function of propagation time inside the PC waveguide. The experimental setup is basically a Michelson interferometer illuminated with a broadband, partially coherent source. The waveguide is placed in one arm and a translating mirror in the other. Properties such as group velocity and complex frequency-dependent reflectance \cite{Combrie:2007} can be extracted from the interference pattern. We also present a powerful and recently developed analytic technique: time-frequency reflectance maps (TFRMs)~\cite{Parini:2008}, that can be used to visualize the frequency-dependent impulse response and reveal a number of interesting features such as disorder-induced scattering and facet reflections.
%, and long trapping of the light in the structure.
%Our non-perturbative theory using a fully solvable 3D model with no fit parameters, presented along side several key experiments, brings to the fore an unparalleled and undisputed view into the rich and much misunderstood role of disorder-induced scattering in PC waveguides.

\rev{Our formalism also provides fresh insights into the long standing question of light localization in PC waveguides. It was proposed by John~\cite{John:1984} and Anderson~\cite{Anderson:1985}, that strong localization may be observable in PCs. Although Anderson localization was originally proposed for electrons propagating in a disordered atomic lattice, the phenomena is general to waves in a periodic medium and applicable to PCs. Anderson localization occurs when the mean free path $l$, of a propagating Bloch mode is reduced to the order of the Bloch wave vector $k$, totally disrupting propagation; formally $kl < 1$~\cite{John:1987}.
There have been a number of milestones towards observing Anderson localization in PCs including the observation of localization in a purely random powder \cite{Wiersma:1997}, and localization transverse to the propagation direction \cite{Schwartz:2007,Lahini:2008}. Recently, Topolancik \textit{et al.}\ observed resonances at the band edge of an artificially roughened PC waveguide and argued these were due to localization~\cite{Topolancik:2007}.
In contrast, here we report measurements on structures with only \emph{unavoidable disorder}, and provide
 %Inspired by the analysis of \citet{Vlasov:1999}, we also provide
 deeper insight into localization phenomena and the correct criteria for strong localization.
} %We cite Vlasov in the discussion
% section introduction (end)

\section{Theory} % (fold)
In the spirit of a coupled mode approach,
we introduce ``slowly-varying envelopes,'' $\psi_{\rm f[b]}(x)$ %[$\psi_{\rm b}(x)$]
for the forward [backward] wave, to approximate the solution as
\begin{eqnarray}
  \mathbf E(\R;\omega) \!= \!\mathcal E_0 [\e_k(\R)\,e^{ikx}\,\psi_{\rm f}(x) + \e_k^*(\R)\,e^{-ikx}\,\psi_{\rm b}(x)]
   + \ldots,\
  \label{eqn:waves}
\end{eqnarray}
where `$\ldots$' includes contributions from lossy radiation modes,
$\mathcal E_0$ is an amplitude, $\e_k(\R)$ is the periodic Bloch-mode electric field normalized by $\int_\mathrm{cell} \id\R \, \varepsilon_i(\R) \, |\e_k(\R)|^2 = 1$, and $k$ is the wave vector which implicitly depends on the the angular frequency $\omega$.

We begin with an ideal structure (no disorder), described through the spatially-dependent dielectric constant $\varepsilon_i(\R)$, as schematically illustrated in Figure~\ref{fig:structure}. The structure supports a Bloch waveguide mode (see the solid blue curve in the band structure plot), which exists below the light line (and is thus lossless in the absence of disorder), and which has a group velocity $v_g = \id \omega / \id k$ that tends to zero at the band edge. If the electric field of the ideal mode $\mathbf E_i(\R;\omega)$, is incident on a disordered waveguide described by $\varepsilon(\R) = \varepsilon_i(\R) + \Delta\varepsilon(\R)$, the exact electric field has a well known analytical form:
$
  \mathbf E(\R;\omega) = \mathbf E_i(\R;\omega) + \int \id\R' \,\G(\R,\R';\omega) \cdot [\mathbf E(\R';\omega) \,\Delta\varepsilon(\R')] ,
$
where $\G(\R, \R';\omega)$ is the Green function for the ideal structure. The Green function is obtained from a polarization-dipole solution to Maxwell's electrodynamics equations. By writing $\mathbf E(\R;\omega)$ as in Eq.\,(\ref{eqn:waves}) and by decomposing the Green function as the superposition of an \emph{exact} bound mode contribution~\cite{Hughes:2005} and a radiation modes background summation $\G_\mathrm{rad}(\R,\R';\omega)$, we then derive a pair of coupled-mode equations for the evolution of the envelopes:
\begin{eqnarray}
  v_g \frac{\id\psi_{\rm f}(x)}{\id x} &=& i\, c_{\rm ff}(x)\, \psi_{\rm f}(x) + i\, c_{\rm fb}(x)\, e^{-i2kx}\, \psi_{\rm b}(x) \nonumber \\
    && + i\, c_{\rm fr}(x)\, \psi_{\rm f}(x), \label{eqn:coupled1} \\
  -v_g \frac{\id\psi_{\rm b}(x)}{\id x} &=& i\, c_{\rm bb}(x)\, \psi_{\rm b}(x) +
i\, c_{\rm bf}(x)\, e^{i2kx}\, \psi_{\rm f}(x) \nonumber \\
    && + i\, c_{\rm br}(x)\, \psi_{\rm b}(x). \label{eqn:coupled2}
\end{eqnarray}
The coupling coefficients can be physically interpreted as $c_{\rm ff} = c_{\rm bb}$ driving scattering from a mode into itself, $c_\mathrm{bf} = c_\mathrm{fb}^*$ driving scattering into the counter-propagating mode, and $c_{\rm fr}$ and $c_{\rm br}$ driving scattering from the waveguide mode into radiation modes above the light line. We have
\begin{eqnarray}
	c_{\rm ff}(x) &=&  \frac{\omega a}{2} \iint\!\! \id y\, \id z\, \e^*_k(\R) \cdot \e_k(\R) \, \Delta\varepsilon(\R), \\
	c_{\rm bf}(x) &=& \frac{\omega a}{2} \iint\!\! \id y\, \id z\, \e_k(\R) \cdot \e_k(\R) \, \Delta\varepsilon(\R), \\
	c_{n \mathrm r}(x) &=&  \frac{\omega a}{2} \iiint\!\! \id y\, \id z\, \id\mathbf r'\, \Delta\varepsilon(\R) \, \tilde\e_{n,k}^*(\R) \cdot \G_\mathrm{rad}(\R,\R'; \omega) \nonumber\\
	    && \quad \quad \quad {} \cdot \tilde\e_{n,k}(\mathbf r') \, \Delta\varepsilon(\R'), \quad n=\mathrm{f, b},
\end{eqnarray}
where $\tilde\e_{\mathrm f, k}(\R) = \tilde\e_{\mathrm b, k}^*(\R) = \e_k(\R) \, e^{ikx}$, and
$a$ is the periodicity of the PC. The most interesting frequency response of the system is dominated by scattering between wave\-guide modes, through $v_g(\omega)$; while radiative scattering merely leaks energy from the waveguide and is a much smaller effect in the slow-light regime as shown in previous studies.
%We calculate effective radiation coefficients $c_{n \mathrm r}^\mathrm{eff}$ using our incoherent theory, and then solve the coupled mode equations numerically for disordered waveguides ``instances.''
\rev{Importantly, this theory thoroughly extends previous formalisms, e.g.~\cite{Hughes:2005}, to account for $i$) multiple scattering events, $ii$) coherent scattering, and $iii$) a diminished Bloch mode amplitude. In addition,
we  solve the coupled mode equations numerically for disordered waveguides ``instances,''
in a fully self-consistent way.}
%In what follows, we
%will use no fit parameters at all, and all parameters are obtained experiment in a self-consistent way.}
% section theory (end)

\section{Experimental Device} % (fold)
\label{sec:experimental_device}
Except where stated otherwise, the experimental device is a W1.1 PC waveguide fabricated from GaAs as schematically shown in Figure~\ref{fig:structure}. The width of the waveguide is $1.1\sqrt3\, a$, the pitch is $a = 400\nm$, the thickness is $265\nm$, the hole radius is $R = 0.27\,a$, and the length is 250\um.
%Recent experiments with GaAs cavities have demonstrated that
The fabrication quality of our GaAs devices is comparable to state-of-the-art silicon processes~\cite{Combrie:2008}.
The samples were analyzed using the high resolution SEM and the image processing technique of \citet{Skorobogatiy:2005}. The disorder was found to be well described by small perturbations of the radius around the hole perimeter $\Delta R(\phi_\alpha)$, that follow the distribution
$
  \langle \Delta R(\phi_\alpha) \Delta R(\phi'_{\alpha'}) \rangle = \sigma^2 e^{R |\phi_\alpha - \phi_{\alpha'}'| / l_p} \delta_{\alpha,\alpha'},
$
where $\alpha$ indexes the holes, and $\phi_\alpha$ is the angular coordinate of the point. The RMS roughness $\sigma$, and correlation length $l_p$, are estimated to be $3$ and $40\nm$ respectively and these values are used in our calculations.
% section experimental_device (end)

\section{Transmission spectra} % (fold)
\begin{figure}[b]
	\centering
	\includegraphics[width=3.375in]{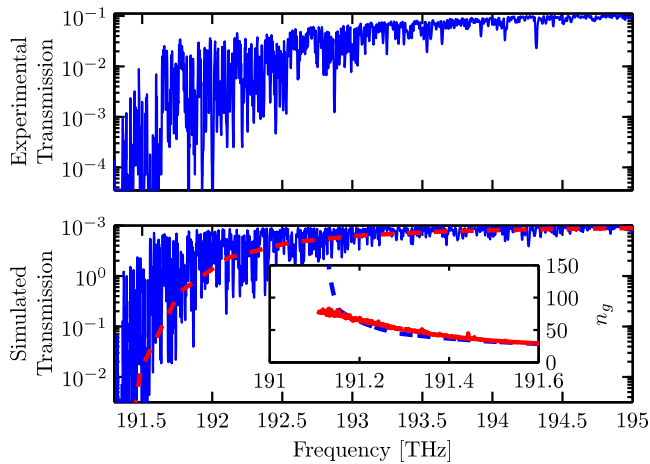}
\vspace{-0.6cm}
	\caption[]{\label{fig:spect}
		Top: Experimental transmission spectra for a 1.5\mm W1 wave\-guide showing resonances near the band edge.
		Bottom: Theoretical transmission spectra calculated using the incoherent (dashed, red) and coherent (solid, blue) scattering theories.
		Inset: Ideal group index $n_g=c/v_g$ (smooth, blue) compared with an estimate of the effective value due to disorder (red, rough). \rev{The effective group velocity is softened due to disorder and does not diverge like the ideal value.}
		\rev{The experimental spectrum is shifted by $1\THz$ to account for uncertainties with the actual fabricated slab thickness.}
	}
\end{figure}
An experimental transmission spectrum is shown in the top plot of Figure~\ref{fig:spect} for a 1.5\mm W1 waveguide of a different design. Approaching the band edge, the transmission rolls-off approximately with $1/v_g^2$, however there are numerous sharp resonances where the transmission varies by orders of magnitude. Two theoretical models for this waveguide are shown in the bottom plot. The previous incoherent loss calculation \cite{Hughes:2005} (dashed, red) captures the approximate $1/v_g^2$ roll-off but does not explain the resonances. In contrast, the new coherent loss calculation presented in this paper (solid, blue) reproduces them since it accounts for multiple scattering events which are necessary to build up Fabry-P\'erot-like resonance between disorder sites. For reference the group index is show in the inset (smooth,~blue).

While we obtain the general trends of the experiments, over more than three orders of magnitude (and
 without any fit parameters), we have not included the disorder-induced broadening of the slow-light-regime band structure which alters the effective group velocity~\cite{Pedersen:2008} and softens the roll-off. \rev{From a simple perspective, disorder adds or removes dielectric from an unperturbed (to first order) Bloch mode, causing a local frequency shift of the band structure. A propagating mode at fixed frequency thus has a locally-varying group velocity due to disorder shifting the waveguide band in frequency. From perturbative calculations with identical disorder statistics, we estimate that the group velocity will be noticeably altered from the ideal value for $v_g \lesssim c/35$ and will have a minimum of around $c/80$ at the band edge, as shown in the inset of Figure~\ref{fig:spect} (rough, red).} For different structures, this minimum $v_g$ will vary. Although these findings are not important for the reflectance maps analyzed below, our calculations are broadly consistent with our own experiments, those reported by \citet{Engelen:2008}, and theoretical analysis of PC coupled-cavity structures~\cite{Fussell:2008}.
% Thales experimental saturation of group velocity is around n_g=20--40.
% section transmission spectra (end)

\section{Time-Frequency Reflectance Maps} % (fold)
TFRMs are intensity plots of the reflected signal as a function of time (horizontal axis) when the structure is excited with a narrow-band pulse centred at some frequency of interest (vertical axis). The map is generated using the complex reflectance of the waveguide which, for physical samples, is deduced from a single set of OLCR data or, for simulated structure, is calculated directly~\cite{Parini:2008}.
\begin{figure}[b]
  \centering
  \vspace{-0.4cm}
  \includegraphics{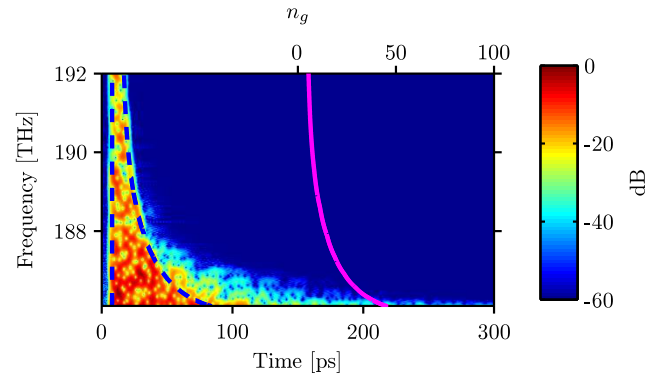}
        \vspace{-0.5cm}
  \caption[]{\label{fig:tfrmW1_1_random1_Rad_NoFacet_NoDomega} A TFRM for a 250\um simulated disordered waveguide with perfectly transmissive facets showing the strength of the back reflection in dB. The left blue dashed line indicates when the pules was injected and the blue dashed curve is the expected round trip time in an ideal structure. The magenta line shows the group index $n_g=c/v_g$ on the top scale.}
       % \vspace{-0.5cm}
\end{figure}

A simulated TFRM is shown in Figure~\ref{fig:tfrmW1_1_random1_Rad_NoFacet_NoDomega} for a wave\-guide with \emph{perfectly transmissive facets}. The dashed blue lines indicate the time the pulse is injected and the round-trip time. The solid magenta line indicates the ideal group index $n_g = c/v_g$ for comparison. Away from the band edge, at higher frequencies, the back reflections are small and are confined between the two time limits, indicating that only single scattering events are occurring. Approaching the band edge, for $v_g<c/20$, scattering becomes significant with strong back reflections and multiple scattering events clearly evidenced by the \emph{hot spots} and the continued reflections after the time for one round trip. This agrees well with \citet{Engelen:2008} who observed total disruption of propagation for $v_g \lesssim c/30$.

\begin{figure}
	\centering
	\includegraphics{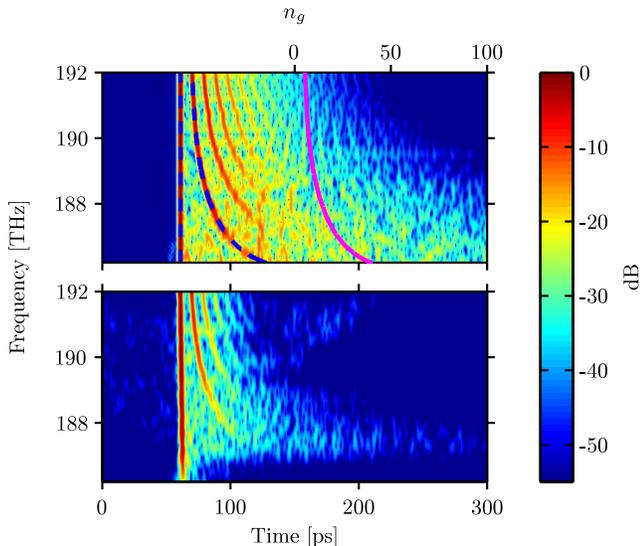} % Experimental data set 11
        \vspace{-0.5cm}
	\caption{\label{fig:tfrmcomparison} Comparison of simulated (top) and experimental (bottom) TFRMs for a 250\um waveguide with a partially reflective front facet and a strongly reflective back facet. The simulation uses the same disorder configuration as in Figure~\ref{fig:tfrmW1_1_random1_Rad_NoFacet_NoDomega}.}
        \vspace{-0.2cm}
\end{figure}

Experimental samples have more complicated TFRMs due to reflections from the sample facets. Figure~\ref{fig:tfrmcomparison} compares simulated (top) and experimental (bottom) TFRMs for the same device geometry. The multiple facet reflections \footnote{We use 50 and 100\% reflectances for the front and back facets respectively to match the sample properties.} are clearly visible and the time for a round trip lengthens as the group velocity decrease. At low group velocity, the pulse is washed out by strong multiple scattering events, making transmission of signals difficult. There is an excellent correspondence between the measured and simulated maps. Since the simulations do not incorporate details of the experimental setup beyond the sample, they tend to give richer features than the experimental maps. Nevertheless, there is an {\em excellent correspondence between the measured and simulated maps}, and this agreement has been found for a number of different sample lengths and facet reflections.
%section TFRM (end)

\section{Localization} % (fold)
Both our measured and simulated transmission spectra (Figure~\ref{fig:structure}) exhibits sharp resonances near the band edge, similar to those reported in \cite{Topolancik:2007}. These features can also be resolved in high resolution TFRMs. To rigorously address the question of whether these features are indicative of localization, a localization length $l$ can be defined as
$
	l^{-1} = -\langle\ln T\rangle / L
$
where $T$ is the transmitted power and $L$ is the sample length~\cite{Vlasov:1999}. For the experimental structure at $k = 0.45\times 2\pi/a$ where $v_g = c / 45$, the localization length is calculated to be $\sim 100\um$, far from the criteria for strong localization. Experimental imaging of light leaking from the waveguide plane confirms that these features are distributed over a large number of waveguide periods. Thus these features are better described as Fabry-P\'erot-like resonances between scattering sites and not localization, in agreement with the interpretation of \citet{Vlasov:1999}. This in no way alters the fact that backscattering in slow light modes leads to highly disordered propagation and low transmission,
which has important implications for  fundamental optical physics and slow light applications.
%with slow light.
% (end)

\section{Conclusion} % (fold)
%We have presented a new coherent loss formalism and matching experiments that allows us to calculate and directly compare the properties of naturally disordered PC waveguides by considering the full 3D structure. In particular, we capture the behavior of a slow-light regime with narrow-band resonances superimposed on the average $1/v_g^2$ roll-off of the incoherent theory. Our simulated waveguides show excellent qualitative agreement with measurements on GaAs waveguides, as demonstrated by comparing TFRMs. We also determine the localization length near the band edge of a PC waveguide and show that while propagation is highly disordered, strong localization is not occurring.
%
\rev{By  matching several carefully-designed light propagation experiments with a new non-perturbative, coherent scattering formalism, we have %been able
 demonstrated and explained the  crossover from nominal light propagation, with a well
 defined group velocity, into a naturally disordered regime
 that is  dominated by coherent scattering and light localization.}
% PC waveguides by considering the full 3D structure.}
Our simulated waveguides show excellent qualitative agreement with measurements on GaAs waveguides, as demonstrated by comparing the rich features of the TFRMs. We also determine the localization length near the band edge of a PC waveguide and show that while propagation is highly disordered, \
%rev{and rich with non-perturbative extrinsic scattering features},
strong localization is not occurring.

We thank Lora Ramunno, Jeff Young, and John Sipe  for useful discussions. This work was supported by the National Sciences and Engineering Research Council of Canada, the Canadian Foundation for Innovation,
 and
 %S. Combri\'e, A. De Rossi, and Q. Tran thank
 the EC project ``GOSPEL'', grant no. 219299.
% (end)

% Choose one of the below bibstyles or comment both to use the revtex4 default
% \bibliographystyle{myapsrev} % To display highly condensed bibliography (and italicised et al)
% \bibliographystyle{plainnat} % To display all bibliographic information
% \bibliography{Masters}

% \vspace{12pt}
% \noindent\makebox[0pt][l]{\rule{\columnwidth}{1pt}}\rule[2pt]{\columnwidth}{1pt}
% \section{Discussion} % (fold)
% \label{sec:discussion}

% (end)

\end{document}